\documentclass[aps,superscriptaddress]{revtex4}
\usepackage[colorlinks=true, pdfstartview=FitV, linkcolor=blue, citecolor=red, urlcolor=magenta, breaklinks=true]{hyperref}
\usepackage{graphicx}  
\usepackage{amsfonts}
\usepackage{amssymb,amsmath,epsfig}

\begin{document}

\title{Hawking radiation and entropy of a BTZ black hole with minimum length}

\author{M. A. Anacleto}\email{anacleto@df.ufcg.edu.br}
\affiliation{Departamento de F\'{\i}sica, Universidade Federal de Campina Grande
Caixa Postal 10071, 58429-900 Campina Grande, Para\'{\i}ba, Brazil}

\author{F. A. Brito}\email{fabrito@df.ufcg.edu.br}
\affiliation{Departamento de F\'{\i}sica, Universidade Federal de Campina Grande
Caixa Postal 10071, 58429-900 Campina Grande, Para\'{\i}ba, Brazil}
\affiliation{Departamento de F\'isica, Universidade Federal da Para\'iba, 
Caixa Postal 5008, 58051-970 Jo\~ao Pessoa, Para\'iba, Brazil}

\author{E. Passos}\email{passos@df.ufcg.edu.br}
\affiliation{Departamento de F\'{\i}sica, Universidade Federal de Campina Grande
Caixa Postal 10071, 58429-900 Campina Grande, Para\'{\i}ba, Brazil}

\author{Jos\'e L. Paulino}\email{leonardopaulino$_$pl@hotmail.com}
\affiliation{Departamento de F\'isica, Universidade Federal da Para\'iba, 
Caixa Postal 5008, 58051-970 Jo\~ao Pessoa, Para\'iba, Brazil}

\author{A. T. N. Silva}\email{andersont.fisica@gmail.com}
\affiliation{Departamento de Física, Universidade Federal de Sergipe,\\
 49100-000 Aracaju, Sergipe, Brazil}

\author{J. Spinelly}\email{jspinelly@uepb.edu.br}
\affiliation{Departamento de F\'{\i}sica-CCT, Universidade Estadual da Para\'{\i}ba,\\
Juv\^encio Arruda S/N, Campina Grande, Para\'iba, Brazil}

\begin{abstract} 
In this paper we consider a BTZ black hole with minimum length which has been introduced through the probability density of the ground state of the hydrogen atom. We analyzed the effect of the minimum length by calculating the thermodynamic quantities such as temperature and entropy and verified the stability of the black hole by computing the specific heat capacity.
\end{abstract}

\maketitle
\pretolerance10000

\section{Introduction}
Modified gravity theories such as those embedded into string theory as low energy effective theories~\cite{Amati, Konishi, Kato}, noncommutative geometry~\cite{Girelli} and loop quantum gravity were proposed in an attempt to construct a self-consistent quantum theory of gravity~\cite{Garay}.
These theories have as characteristics the existence of a minimum length in the order of the Planck scale. 
This minimum length leads to a modification of the Heisenberg uncertainty principle and has been called the generalized Heisenberg uncertainty principle (GUP)~\cite{Casadio:2014pia,Tawfik:2015pqa,Nozari:2008gp}.
An important consequence of the GUP in the study of black hole thermodynamics is the existence of a minimum mass, which has the effect of stopping the evaporation of the black hole~\cite{Adler:2001vs,Banerjee:2010sd,Gangopadhyay:2013ofa,Dutta:2014yna,Tawfik:2013uza, Anacleto:2015rlz,Anacleto:2015kca,Gomes:2018oyd,Kuntz:2019gka,Chen:2002tu,Chen:2013pra,Chen:2013tha,Feng:2015jlj,
Matsuno:2021pvz,Ali:2019xxd}.
Moreover, it has been found logarithmic corrections for the calculation of entropy due to GUP~\cite{Anacleto:2015mma,Anacleto:2015awa,Gangopadhyay:2014wja,Majumder:2012rtc,Bargueno:2015tea,
Ovgun:2015box,Sadeghi:2016xym,Maluf:2018lyu,Mele:2021hro,Haldar:2018zyv,Fu:2020ckx,Giardino:2020myz,Anacleto:2021nhm}. Also, by considering the GUP, the thermodynamics of black holes was investigated by means of several approaches~\cite{Ali:2012mt,Majumder:2011bv,Bina:2010ir,Xiang:2009yq,Kim:2007hf, Li:2016mwq,Ovgun:2017hje,Hendi:2016hbe,Meitei:2018mgo,Gecim:2017nbh,Gecim:2017zid,Gecim:2017lxx,Casadio:2017sze,Alonso-Serrano:2021eju}.
Besides, the entropy of several types of black holes has also been analyzed by considering a state density corrected through the GUP~\cite{Sakalli:2016mnk,Faizal:2014tea,SG2015,ADV}, and thus the divergences that could appear in the brick wall model are eliminated without the need for a cut-off. 

In the last years a large number of works have been done by considering  the metric  of  black holes deformed by the presence of noncommutativity.
In~\cite{NSS}, the authors introduced noncommutativity in the study of black holes by considering that the effects of noncommutativity can be implemented by modifying the source of matter in Einstein's equations.
In this way, they modified the energy-momentum tensor on the right hand side of Einstein's equations and kept the left hand side of these equations unchanged. Thus, they considered a Gaussian distribution function rather than a $ \delta $ function for the mass density as a gravitational source.
Therefore, by solving the Einstein equations with these modifications, was determined the metric of a deformed Schwarzschild black hole due to noncommutativity. 
In addition, many papers have appeared in the literature suggesting different ways of introducing 
noncommutativity into Ba\~nados-Teitelboim-Zanelli (BTZ) black holes~\cite{BTZ,Nicoline2009,Banados2001,Kim:2007nx,Chang-Young:2008zbi,Eune:2013qs,Anacleto:2014cga,Sadeghi:2015nzp,Hendi:2015wxa,Hussein,Gecim:2020zcb}.
{Furthermore, the thermodynamic properties of noncommutative BTZ and noncommutative Schwarzschild black holes have been explored via Lorentzian distribution with GUP and logarithmic corrections for entropy have been obtained due to the noncommutativity effect~\cite{Anacleto:2020efy,Anacleto:2020zfh}. 
Also an analysis of the quasinormal modes and shadow radius of the noncommutative Schwarzschild black hole introduced by means of a Lorentzian distribution has been performed in~\cite{Campos:2021sff,Zeng:2021dlj}.}
In~\cite{Miao:2016ipk,Liu:2015dfg}, by applying a smeared mass distribution, the authors investigated the thermodynamic properties of a Schwarzschild-AdS  black hole where the mass density was considered to be the probability density of the ground state of the hydrogen atom.

In the present work, inspired by all the aforementioned results, we consider a BTZ black hole modified by the presence of a minimum length. 
The effect of the minimum length in the BTZ metric is obtained by replacing the mass density by the probability density of the ground state of the hydrogen atom in two spatial dimensions.
Thus, we compute the Hawking temperature and the entropy of the modified black hole. 
In addition, we also determined the specific heat capacity in order to explore the stability of the modified BTZ black hole due to the minimum length. 
The minimum length effect can have interesting implications in condensed matter physics, for example, in the investigation of the effective theory describing the spin-orbit interaction, Landau levels, quantum Hall effect, as well as in superconductivity --- see for instance~\cite{Das:2011tq,Khan:2022yfm}.

The paper is organized as follows. In Sec.~\ref{btzml} we consider minimum length corrections for the BTZ black hole metric implemented via the probability density of the ground state of the hydrogen atom in two dimensions, smeared distribution on a collapsed shell and Lorentzian-type distribution as the mass density for the BTZ black hole. 
In Sec.~\ref{conc} we make our final considerations.

\section{BTZ Black Hole with minimum length}
\label{btzml}

In this section we incorporate minimum length corrections for the BTZ metric by modifying the mass density of the black hole through the probability density of hydrogen in two dimensions. 
Thus, we will investigate Hawking radiation, entropy and the stability of the BTZ black hole.

\subsection{Mass Density - Probability Density }

In our model,  based on the work~\cite{Miao:2016ipk}, we consider the probability density of the ground state of the hydrogen atom in two dimensions~\cite{Yang1991} as the mass density for the BTZ black hole
\begin{eqnarray}
\rho(r) = \frac{M}{a^{2}\pi} \exp\left(\frac{-4r}{a}\right),
\label{a1}
\end{eqnarray}
where $M$ is the total mass of the BTZ black hole and $a$ is the minimum length parameter in the model.

{In order to solve Einstein's equations, in addition to knowing the energy-momentum tensor, we must write the metric in a way that suits the symmetry of the distribution. Thus, in the present case, the line element describing the three-dimensional spacetime must be of the type.
\begin{eqnarray}
ds^2=-A(r)dt^2 + B(r)dr^2 + r^2d\phi^2.
\end{eqnarray}
Therefore, taking this into account and using the fact that due to the conservation law $\nabla_{\nu}T^{\mu\nu}=0$, 
the energy-momentum tensor is given by $ T^{\mu}_{\nu}=diag(-\rho, p_r,p_{\perp}) $, where radial and tangential pressure are described by
${p}_r=-{\rho}$ and ${p}_\bot=-{\rho}rd{\rho}/dr$, respectively, follows from Einstein's equations that
\begin{equation}
\frac{1}{2B^2r}\frac{dB}{dr}-\Lambda =2\pi \rho , 
\label{Y12}
\end{equation}
and
\begin{equation}
\frac{1}{2BAr}\frac{dA}{dr}+\Lambda =-2\pi \rho \ .
\label{Y13}
\end{equation}
By solving this system of equations, we obtain
\begin{eqnarray}
A=\frac{\alpha}{B},
\end{eqnarray}
and
\begin{eqnarray}
B=\frac{1}{-2\mathcal{M}-\Lambda r^2+\beta} =\frac{1}{f(r)}\ ,
\end{eqnarray}
where $ \alpha $ and $ \beta $ are constants and $ \mathcal{M}  $ is the spread mass distribution contained in a region of radius $r$, given by
\begin{eqnarray}
\mathcal{M}(r) =   \int^{r}_{0} \rho(r)2\pi r dr = M\left[1-\frac{(4r+a)}{a}\exp\left(\frac{-4r}{a}\right)\right].
\label{a2}
\end{eqnarray}
Now, making $ \alpha=1 $, $ \beta=M $ and $ \Lambda=-1/l^2 $, we obtain
\begin{eqnarray}
ds^2=-f(r)dt^2+f^{-1}(r)dr^2+r^2d\varphi^2,
\label{btzcm}
\end{eqnarray}
where
\begin{eqnarray}
f(r)=-2\mathcal{M}(r)+\frac{r^2}{l^2}+M=-M  +\frac{r^2}{l^2}+ \left[\frac{(8Mr+2a)}{a}\exp\left(\frac{-4r}{a}\right)\right].
\label{a05}
\end{eqnarray}
Note that in the limit $r/a\rightarrow\infty$,  we recover the nonrotating BTZ black hole metric. 

The event horizon of the BTZ black hole is obtained when we make $ f(r) = 0 $. 
Thus, we have
\begin{eqnarray}
r_H^2=r_h^2\left[1-2\left(\frac{l^2}{r^2_h}+ \frac{4r_H}{a} \right)e^{-4 r_H/a} \right].
\end{eqnarray} 
where $ r_h=\sqrt{Ml^2} $ is the event horizon for the nonrotating BTZ black hole in the absence of the minimum length. 
Now, solving the above expression by iteration and keeping terms up to first order in $ e^{-4 r_h/a} $,  we obtain
\begin{eqnarray}
r_H=r_h\left[1-2\left(\frac{l^2}{r^2_h}+ \frac{4r_H}{a} \right)e^{-4 r_H/a} \right]^{1/2}
=r_h\left[1-\left(\frac{l^2}{r^2_h}+ \frac{4r_h}{a} \right)e^{-4 r_h/a} \right] +{\cal O}\left(e^{-4r_h/a}\right)^2.
\end{eqnarray}

The Hawking temperature of the nonrotating BTZ black hole  with minimum length  is given by
\begin{eqnarray}
T_H=\frac{1}{4\pi}\frac{\partial f}{\partial r}{\Big|}_{r_H}=\frac{r_H}{2\pi l^2}\left[1-\frac{16Ml^2}{a^2}e^{-4r_H/a}\right].
\label{je1}
\end{eqnarray}
In terms of $ r_h $ we can write the Hawking temperature as follows 
\begin{eqnarray}
\label{TH}
T_H=\frac{r_h}{2\pi l^2} -\frac{1}{2\pi l^2}\left[\frac{l^2}{r_h}+\frac{4r^2_h}{a} +\frac{16r^3_h}{a^2}\right]e^{-4r_h/a} 
+{\cal O}\left(e^{-4r_h/a}\right)^2.
\end{eqnarray}
Note that the first term is the Hawking temperature of the BTZ black hole ($ T_h=r_h/2\pi l^2 $) and  the second term is the correction due to the minimum length.

From the condition $f(r)=0$, we can obtain the mass of the BTZ black hole which is given by 
\begin{eqnarray}
M=\frac{r^2_H}{l^2}\left[1-2\left( \frac{l^2}{r^2_h}+\frac{4r_H}{a}\right)\exp\left(-\frac{4r_H}{a}\right)\right]^{-1},
\label{massa}
\end{eqnarray}
or in terms of $ r_h $, we have
\begin{eqnarray}
M=\frac{r^2_h}{l^2} + \left(e^{-4r_h/a}\right)^{2}.
\label{masrh}
\end{eqnarray}
Next, in order to compute the entropy of a BTZ black hole with minimum length we consider the following equation:
\begin{eqnarray}
S=\int\frac{dM}{T_H}=\int\frac{1}{T_H}\frac{\partial M}{\partial r_h}dr_h,
\label{a9}
\end{eqnarray} 
By replacing (\ref{TH}) and (\ref{masrh}) in (\ref{a9}), we find
\begin{eqnarray}
S&=&\int \frac{2\pi l^2}{r_h}\left[1+ \left(\frac{l^2}{r^2_h}+\frac{4r_h}{a} +\frac{16r^2_h}{a^2}\right)e^{-4r_h/a} + \cdots\right]\left(\frac{2r_h}{l^2} \right) dr_h,
\\
&=&4\pi r_h - \frac{4l^2}{a} {\mbox Ei}\left(\frac{-4r_h}{a} \right) -\left[\frac{3a}{4} +\frac{l^2}{r_h} + 3r_h +\frac{4r^2_h}{a} \right]e^{-4r_h/a} +\cdots .
\end{eqnarray}

Now we calculate the correction of the specific heat capacity  which is given by
\begin{eqnarray}
\label{shcg}
C&=&\frac{\partial M}{\partial T}=\frac{\partial M}{\partial r_h}\left(\frac{\partial T}{\partial r_h}\right)^{-1},
\\
&=& 4\pi r_h \left[ 1+ \left( R_2 -\frac{4R_1}{a}\right) e^{-4r_h/a}\right],
\end{eqnarray}
where
\begin{eqnarray}
&& R_1=\frac{l^2}{r_h} +\frac{4r^2_h}{a} + \frac{16 r^3_h}{a^2},
\\
&& R_2=-\frac{l^2}{r^2_h} +\frac{8r_h}{a} + \frac{48 r^2_h}{a^2}.
\end{eqnarray}
{Note that, considering the dominant term in (\ref{shcg}) in the limit of $a\ll 1$, we have
\begin{eqnarray}
\label{Cs}
 C\approx 4\pi r_h \left[ 1 -\frac{64 r^3_h}{a^3} e^{-4r_h/a}\right].
\end{eqnarray}
Hence the condition that cancels the specific heat capacity is given by
\begin{eqnarray}
\label{condcs}
\frac{ r^3_h}{ e^{4r_h/a}}=\left(\frac{a}{4}\right)^3.
\end{eqnarray}
Now, we apply the above condition to the Hawking temperature expression (\ref{TH})  and thus obtain
\begin{eqnarray}
\label{Tmax}
T_{max}\approx \frac{1}{2\pi l^2}\left( r_h -\frac{a}{4} \right)=T_H-\frac{a}{8\pi l^2}, 
\end{eqnarray}
{which is the maximum temperature of the remnant of the black hole. 
On the other hand, for $r_h/a\ll 1$, from condition (\ref{condcs}), we find $r_h=3a/8$. 
Thus, the result in (\ref{Tmax}) becomes
\begin{eqnarray}
{T_{max}\approx\frac{a}{16\pi l^2},}
\end{eqnarray}
with a minimum mass given by
\begin{eqnarray}
M_{min}=\frac{9a^2}{64l^2}.
\end{eqnarray}
\begin{figure}[htb]
\includegraphics[scale=0.45]{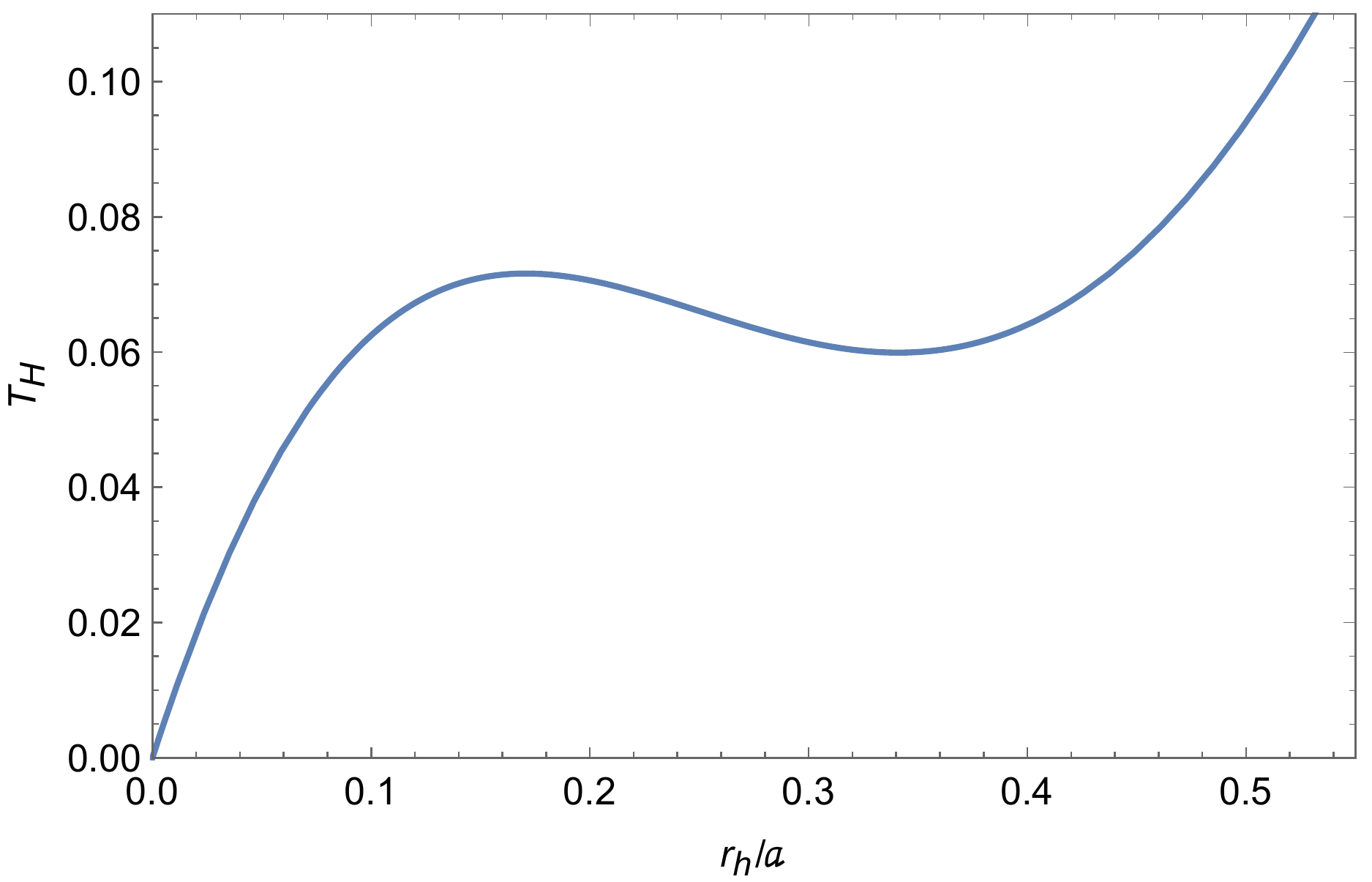}
\caption{The Hawking temperature $ T_H \rightarrow 2\pi l^2 T_H/a$ in function of $ r_h/a $.
Note that the temperature reaches a maximum value $ T_{max} $, and then decreases to zero as
$ r_h/a\rightarrow 0 $.}
\label{thrm}
\end{figure} 
However, as shown in Fig.~\ref{thrm} the remnant reaches a maximum temperature before going to zero ($T_{rem}\rightarrow 0$ for $ r_h\ll a $).}

{In Fig.~\ref{cs1} we show the behavior of the specific heat capacity. 
The graph shows that, for $ a=0.35 $, the specific heat capacity goes to zero at two points and between them there is a non-physical region.}

\begin{figure}[htb]
\includegraphics[scale=1.0]{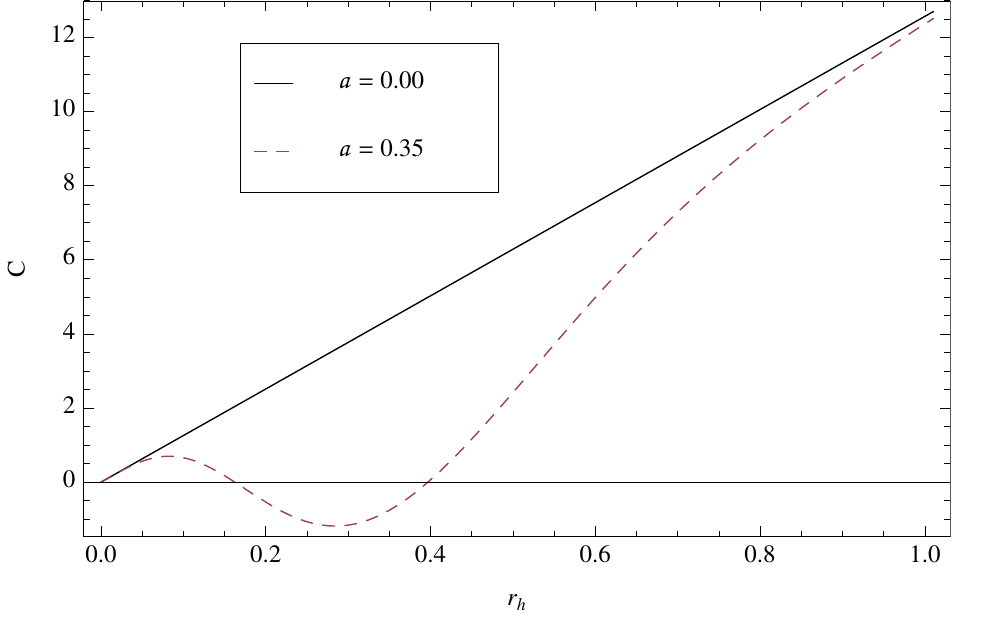}
\caption{Specific heat capacity in function of $ r_h $ (Eq. (\ref{Cs})). The solid line represents the BTZ black hole 
and the dashed line corresponds to the BTZ black hole with minimum length $ a=0.35 $.}
\label{cs1}
\end{figure}

For the rotating black hole with minimum length the line element is given by
\begin{eqnarray}
ds^2=-f(r)dt^2+h^{-1}(r)dr^2-Jdtd\varphi+r^2d\varphi^2,
\label{btzjcm}
\end{eqnarray}
where
\begin{eqnarray}
&&f(r)=-2\mathcal{M}(r)+\frac{r^2}{l^2}+M=-M  +\frac{r^2}{l^2}+ \left[\frac{(8Mr+2a)}{a}\exp\left(\frac{-4r}{a}\right)\right],
\\
&&h(r)=-2\mathcal{M}(r)+\frac{r^2}{l^2}+M+\frac{J^2}{4r^2}=-M  +\frac{r^2}{l^2}+\frac{J^2}{4r^2}+ \left[\frac{(8Mr+2a)}{a}\exp\left(\frac{-4r}{a}\right)\right].
\end{eqnarray}
To determine the Hawking temperature we apply the equation
\begin{eqnarray}
T_H=\frac{\kappa}{2\pi},
\end{eqnarray}
where
\begin{eqnarray}
\kappa^2=\left.-\frac{1}{2}\nabla_{\mu}\chi_{\nu}\nabla^{\mu}\chi^{\nu}\right|_{r=r_+},
\label{grav}
\end{eqnarray}
is the surface gravity, where$\chi^{\mu}=\left(1,0,\Omega\right)$ is the killing field,  being 
\begin{eqnarray}
\Omega\equiv -\left.\frac{g_{t\varphi}}{g_{\varphi \varphi}}\right|_{r=r_+},
\end{eqnarray}
the angular velocity in the event horizon, $r_+$. 

Thus, we can compute the Hawking temperature using the following formula:
\begin{eqnarray}
T_H= \frac{1}{2\pi}\left\{\frac{h}{4fr^2+J^2}\left[r^2\left(\frac{df}{dr}\right)^2-2J\Omega r\frac{df}{dr}-4\Omega^2r^2f\right]\right\}^{1/2}_{r=r_+}\ ,
\end{eqnarray} 
where $\Omega={J}/{2 r^2_{+}}$. 
Hence the Hawking temperature for the rotating BTZ black hole with minimum length is given by
\begin{eqnarray}
T_H&=&\frac{1}{2\pi l^2r_{+}^2}\left\{\frac{ 256 M^2l^4r_+^6}{a^4}\exp\left(-\frac{8r_+}{a}\right) \right. \nonumber\\ 
&&+ M\left[\left(-\frac{32l^2r_{+}^6}{a^2}+\frac{8J^2l^4r_{+}^2}{a^2}-\frac{2J^2l^4r_{+}}{a}-\frac{J^2l^4}{2}\right)\exp\left(-\frac{4r_+}{a}\right) \right. \nonumber \\
&&\left.\left.+\frac{J^2l^4}{4}\right]+r_+^6-\frac{3J^2l^2r_+^2}{4}\right\}^{1/2} \ ,
\label{je2}
\end{eqnarray}
where
\begin{eqnarray}
M=\left(\frac{r^2_+}{l^2} +\frac{J^2}{4r^2_+}\right) \left[1-2\left( \frac{l^2}{r^2_h}+\frac{4r_+}{a}\right)
\exp\left(-\frac{4r_+}{a}\right)\right]^{-1},
\label{masj}
\end{eqnarray}

\subsection{Smeared Distribution on a Collapsed Shell }
{Here, based on the work~\cite{Miao:2016ipk}, we introduce the effect of a minimum length considering a mass distribution given by}
\begin{eqnarray}
\rho(r)=\frac{100Ma^2_0r^2}{2\pi (a^2_0 + 5r^2)^3},
\end{eqnarray}
where $M$ is the total mass of the BTZ black hole and $a_0$ being the minimum length parameter.
The  smeared mass distribution function can be determined by applying the following relationship:
\begin{eqnarray}
\mathcal{M}_0(r) &=&   \int^{r}_{0} \rho(r)2\pi r dr = \frac{25Ma^2_0 r^4}{(a^3_0 + 5a_0r^2)^2},
\\
&=& M - \frac{2M a^2_0}{5r^2} + {\cal O}(a^4_0).
\label{m01}
\end{eqnarray}
Next, we incorporate the minimum length effect into the metric at the BTZ black hole by applying the above mass to construct the line element given by
\begin{eqnarray}
ds^2=-f(r)dt^2+f^{-1}(r)dr^2+r^2{(d\varphi+N^{\varphi}dt)}^2,
\label{a4}
\end{eqnarray}
where
\begin{eqnarray}
f(r)=-M+\frac{r^2}{l^2}+\frac{J^2}{4r^2} + \frac{4M a^2_0}{5r^2} ,
\label{a5}
\end{eqnarray}
and
\begin{eqnarray}
N^\varphi=-\frac{J}{2r^2}.
\label{a6}
\end{eqnarray}
Note that the parameter $ a_0 $ simulates the effect of an ``effective angular momentum" on the metric.

Let us now consider the BTZ black hole in the nonrotating regime ($ J=0 $). Thus, the metric becomes
\begin{eqnarray}
ds^2=-f(r)dt^2+f^{-1}(r)dr^2+r^2d\varphi^2,
\label{metj0}
\end{eqnarray}
where
\begin{eqnarray}
f(r)=-M+\frac{r^2}{l^2} + \frac{4M a^2_0}{5r^2} .
\label{fmetj0}
\end{eqnarray}
By making $f(r)=0$, we find the event horizons given by
\begin{eqnarray}
&&r^2_{+}=\frac{r^2_h}{2}\left[1 + \sqrt{1-\frac{4\tilde{a}^2}{r^2_h}}   \right]=r^2_h\left( 1-\frac{\tilde{a}^2}{r^2_h} +\cdots\right), 
\quad\Rightarrow \quad r_+=r_h - \frac{\tilde{a}^2}{2r_h}+\cdots,
\\
&&r^2_{-}=\frac{r^2_h}{2}\left[1 - \sqrt{1-\frac{4\tilde{a}^2}{r^2_h}}   \right]=\tilde{a}^2 + \cdots, 
\quad\Rightarrow \quad r_{-}=\tilde{a}+\cdots,
\end{eqnarray}
where  $ r_h=\sqrt{Ml^2} $ and $ \tilde{a}^2=4a^2_0/5 $.

In this case, for the Hawking temperature we obtain
\begin{eqnarray}
T_H=\frac{f^{\prime}(r_+)}{4\pi}=\frac{2r_+}{4\pi l^2}\left(1- \frac{r^2_h \tilde{a}^2}{r^4_+}  \right),
\end{eqnarray}
which in terms of $r_h$ becomes
\begin{eqnarray}
T_H=\frac{r_h}{2\pi l^2}- \frac{3\tilde{a}^2}{4\pi l^2r_h} +\cdots.
\end{eqnarray}
The mass of the BTZ black hole is
\begin{equation}
M=\frac{r^2_+}{l^2}\left( 1+ \frac{r^2_h \tilde{a}^2}{r^4_+}  \right)=\frac{r^2_h}{l^2}+ {\cal O}(\tilde{a}^4) .
\end{equation}
For entropy we obtain
\begin{eqnarray}
S&=&\int\frac{1}{T_H}\frac{\partial M}{\partial r_h}dr_h
=4\pi\int \frac{l^2}{r_h}\left[ 1+\frac{3\tilde{a}^2}{2r^2_h} + \cdots \right]\frac{r_h}{l^2} dr_h
\\
&=& 4\pi r_h - \frac{6\pi\tilde{a}^2}{r_h}  + \cdots .
\label{entropycs}
\end{eqnarray} 

In this case the correction for the specific heat capacity is
\begin{eqnarray}
C&=&\frac{\partial M}{\partial r_+}\left(\frac{\partial T}{\partial r_+}\right)^{-1},
\\
&=& 4\pi r_+ \left(1-\frac{\tilde{a}^2}{r^2_+} +\cdots   \right)
\\
&=&4\pi r_h \left(1-\frac{3\tilde{a}^2}{2r^2_h} +\cdots   \right).
\label{csa0}
\end{eqnarray}
For  $ r_h=\sqrt{3}\tilde{a}/\sqrt{2} $  the specific heat vanishes. Thus, we have a minimum radius given by
\begin{eqnarray}
r_{min}=\sqrt{l^2M_{min}}=\frac{\sqrt{3}\tilde{a}}{\sqrt{2}}=\frac{\sqrt{6}a_0}{\sqrt{5}},
\end{eqnarray}
and a minimum mass given by
\begin{eqnarray}
M_{min}=\frac{3\tilde{a}^2}{2 l^2}=\frac{6 a_0^2}{5 l^2}.
\end{eqnarray}
{In this case, we find $T_{rem}=0$ for the temperature of the black hole remnant.}
\begin{figure}[htb]
\includegraphics[scale=1.0]{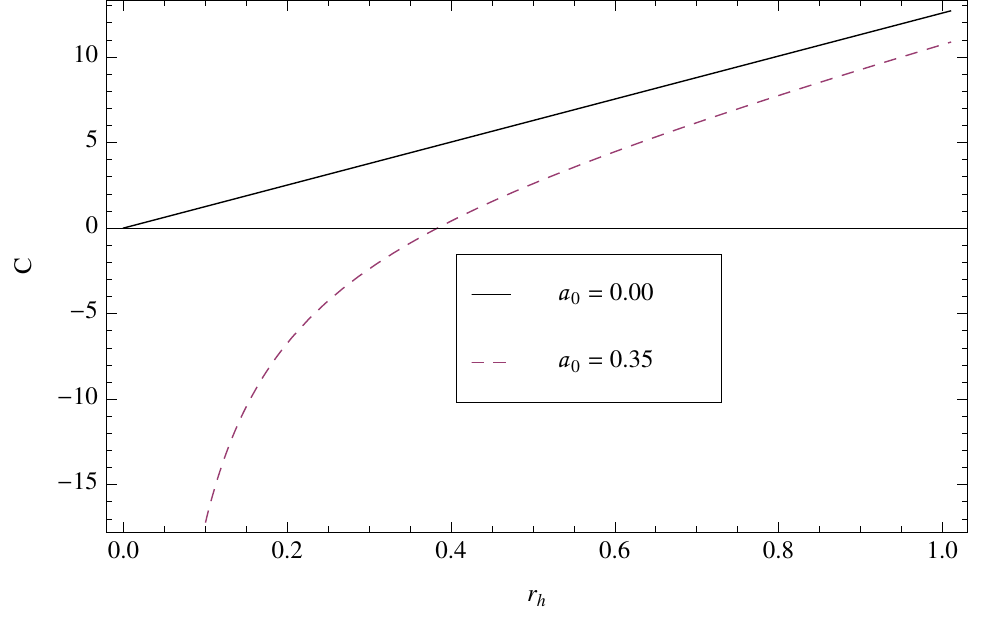}
\caption{Specific heat capacity in function of $ r_h $ (Eq. (\ref{csa0})). The solid line represents the BTZ black hole 
and the dashed line corresponds to the BTZ black hole with minimum length $ a_0=0.35 $.}
\label{cs2}
\end{figure} 

{In Fig.~\ref{cs2} we check the behavior of the specific heat capacity.
The graph shows that, for $a_0=0.35$, the specific heat capacity is stable in the region where $r_h > r_c$ (critical radius) and for $r_h<r_c$ it enters a non-physical region.}

\subsection{Lorentzian-Type Distribution  }

{In this section, based on the work~\cite{Anacleto:2020efy},  we will analyze the effect of minimum length considering the following mass distribution:}
\begin{eqnarray}
\rho(r)=\frac{16Mb}{\pi (4r + b)^3},
\end{eqnarray}
where $ b $ is the minimum length parameter.
Now we determine the mass distribution given by
\begin{eqnarray}
\mathcal{M}_b(r) &=&   \int^{r}_{0} \rho(r)2\pi r dr = \frac{16Mr^2}{(b + 4 r)^2},
\\
&=& M -\frac{M b}{2 r} + \frac{3M b^2}{16r^2} + {\cal O}(b^3).
\label{m1}
\end{eqnarray}
From there, we can get the line element as follows:
\begin{eqnarray}
ds^2=-f(r)dt^2+f^{-1}(r)dr^2+r^2d\varphi^2,
\label{metcs}
\end{eqnarray}
where
\begin{eqnarray}
f(r)=-M +\frac{Mb}{2r}+\frac{r^2}{l^2} - \frac{3M b^2}{16r^2} .
\label{fmetcs}
\end{eqnarray}
Note that by applying the above distribution new terms arise in the metric function (\ref{fmetcs}).
The second term, ${Mb}/{2r}$, is a term of type Schwarzschild and the last term, ${3M b^2}/{16r^2}$, corresponds to an effective angular momentum term.
In our calculations we will consider the situation where $J=0$, so the effect of angular momentum will be simulated by the last term of the metric function above.

The horizons are determined by solving the quartic equation
\begin{eqnarray}
r^4 - l^2 M r^2 - \frac{3l^2 M b^2}{16}+ \frac{l^2 M b}{2} r=0.
\end{eqnarray}
The equation can be rewritten as follows~\cite{Anacleto:2020efy,Visser:2012wu}:
\begin{eqnarray}
\label{eqh}
(r^2-r^2_{+})(r^2-r^2_{-}) + \frac{l^2 Mb}{2}r=0,
\end{eqnarray}
where $r_{+} $ and $ r_{-} $ are given by
\begin{eqnarray}
r^2_{\pm}=\frac{l^2 M}{2}\left(1\pm \sqrt{1+\frac{3b^2}{4l^2 M }}\right)=
\begin{cases}
r^2_{+}=r^2_h+\dfrac{3b^2}{16}+ \cdots,
\\
\\
r^2_{-}=-\dfrac{3b^2}{16}+ \cdots.
\end{cases}
\end{eqnarray}
In order to solve the equation (\ref{eqh}) perturbatively, we write in the form
\begin{eqnarray}
r^2=r^2_{\pm} - \frac{r^2_h br}{2(r^2-r^2_{\mp})},
\end{eqnarray}
where $ r_h=\sqrt{l^2 M} $.
Hence, in the first approximation to the outer radius we have
\begin{eqnarray}
\tilde{r}^2_{+}\approx r^2_{+} + \frac{r^2_h b r_{+}}{2(r^2_{-} - r^2_{+})},
\end{eqnarray}
or
\begin{eqnarray}
\tilde{r}_{+}=r_{+} + \frac{r^2_h b}{4(r^2_{-} - r^2_{+})}+\cdots,
\end{eqnarray}
and for the radius of the inner horizon we find
\begin{eqnarray}
\tilde{r}^2_{-}\approx r^2_{-} - \frac{r^2_h b r_{-}}{2(r^2_{-} - r^2_{+})},
\end{eqnarray}
and so for $ \tilde{r}_{-} $, we have
\begin{eqnarray}
\tilde{r}_{-}=r_{-} - \frac{r^2_h b}{4(r^2_{-} - r^2_{+})}+\cdots.
\end{eqnarray}

The mass of the BTZ black hole is given by
\begin{eqnarray}
M&=&\frac{\tilde{r}_+^{2}}{l^{2}} 
+\frac{\tilde{r}_+ b}{2l^2} - \frac{3b^2}{16l^2} + \cdots.
\end{eqnarray} 
For the Hawking temperature we have
\begin{eqnarray}
\label{ThB}
{\tilde T}_{H}&=&\frac{\tilde{r}_+}{2\pi l^2}
\left(1 - \frac{r^2_h b}{4\tilde{r}^3_+}+\frac{3r^2_hb^2}{16\tilde{r}^4_+}\right),
\\
&=& \dfrac{1}{2\pi l^2}\left(r_h-\dfrac{b}{2}-\dfrac{b^2r_h}{8}+\dfrac{9b^2}{32r_h}  \right).
\label{ThB2}
\end{eqnarray}	
Now by following the same steps done above we can determine the entropy and the specific heat capacity that are respectively given by
\begin{eqnarray}
\hat{S}&=& 4\pi\int \left(1 + \frac{r^2_h b}{4\tilde{r}^3_+}-\frac{3r^2_hb^2}{16\tilde{r}^4_+}\right)
\left(1+\frac{b}{4\tilde{r}_+}\right)d\tilde{r}_+ 
\\
&=&4\pi\tilde{r}_+ + \pi b\ln(\tilde{r}_+) - \frac{\pi r^2_h b}{2\tilde{r}^2_+} 
- \frac{\pi r^2_h b^2}{48\tilde{r}^3_+} +\cdots .
\end{eqnarray}
and
\begin{eqnarray}
\label{ctb}
{C}_{b}=4\pi\tilde{r}_+\left(1 +\frac{b}{4\tilde{r}_+}\right)
\left(1- \frac{b}{2\tilde{r}_+}\right)+\cdots,
\end{eqnarray}
Hence, the specific heat vanishes at the point $\tilde{r}_+=b/2 $ (or $ r_h=\sqrt{6}b/4 $). We soon find that the minimum radius and the minimum mass are
\begin{eqnarray}
r_{bmin}=\sqrt{l^2M_{bmin}}=\frac{\sqrt{6}b}{4},
\end{eqnarray}
and
\begin{eqnarray}
M_{bmin}=\frac{3 b^2}{8l^2}.
\end{eqnarray}
Therefore, with this result the black hole becomes a remnant with the maximum temperature given by
\begin{eqnarray}
T_{bmax}=0.4T_H=\frac{\sqrt{6}b}{20\pi l^2},
\end{eqnarray}
{and for $ b\ll 1 $, we have from Eq. (\ref{ThB2}) $ T_{brem}\rightarrow 0 $.

Note that since the Hawking temperature of the BTZ black hole is directly proportional to the radius of the horizon ($ T_H \sim r_h $), the maximum temperature of the remnant is directly proportional to the minimum length 
($ T_{bmax} \sim b $), and therefore becomes smaller than the Hawking temperature of the BTZ black hole
($T_{bmax} < T_H $)~\cite{Anacleto:2020efy}.  }

{In Fig.~\ref{cs3} we analyzed the stability  of the specific heat capacity.
The graph shows that, for $b=0.35$, the specific heat capacity is stable in the region where $r_h > r_c$ (critical radius) and for $r_h<r_c$ it enters a non-physical region. 
The BTZ black hole with minimum length decreases in size until it reaches a critical radius where it stops evaporating completely and then becomes a black hole remnant.}
\begin{figure}[htb]
\includegraphics[scale=1.0]{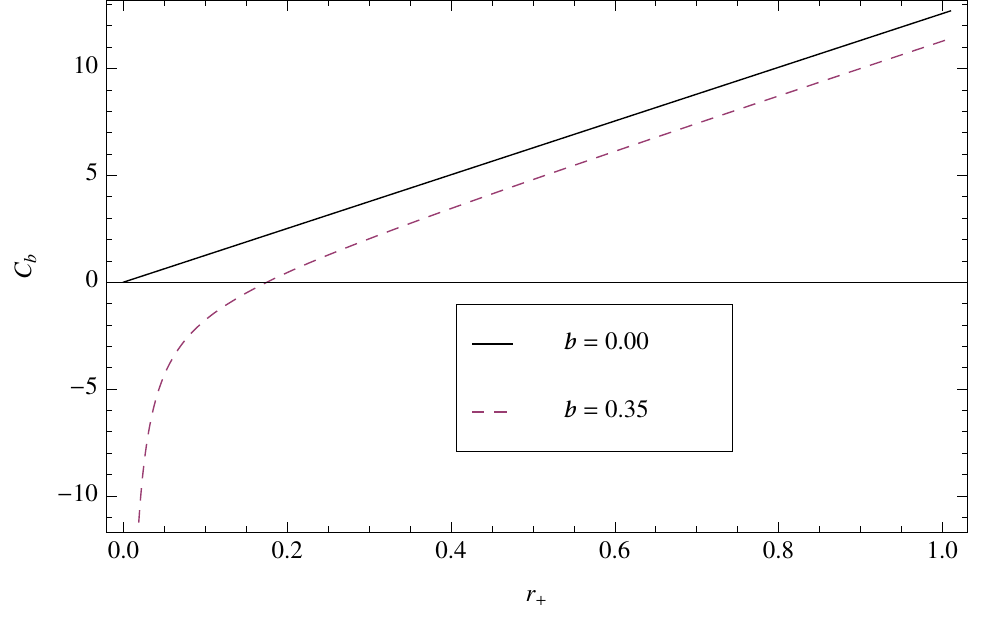}
\caption{Specific heat capacity in function of $ \tilde{r}_+ $ (Eq. (\ref{ctb})). The solid line represents the BTZ black hole 
and the dashed line corresponds to the BTZ black hole with minimum length $ b=0.35 $.}
\label{cs3}
\end{figure} 

 Note also that in the limit of $ M \rightarrow 0 $ ($r_h\rightarrow 0$) we have 
$ \tilde{r}_+=r_b=\sqrt{3}b/4 $.
In this case even in the absence of the mass parameter $M$, we find a non-zero result 
due to the effect of the minimum length for the temperature, entropy 
and specific heat capacity of the BTZ black hole, which are respectively given by
\begin{eqnarray}
&& T_{b}=\frac{r_b}{2\pi l^2}=\frac{\sqrt{3}b}{8\pi l^2},
\\
&& S_b=4\pi r_b + \pi b\ln(r_b) ,
\\
&& C_b=4\pi r_b(1+1/\sqrt{3}) =\pi(1+\sqrt{3})b.
\end{eqnarray}
Furthermore, we verified that even at the zero mass limit the logarithmic correction term is obtained in the calculation of entropy. In addition, as $C_b>0$ the black hole remains stable at this limit.
It is worth mentioning that in~\cite{Singleton:2010gz} a similar form for entropy with logarithmic correction has been found using a phenomenological approach to quantum gravity, and also an analysis when $M \rightarrow 0$ for the Schwarzschild black hole with quantum corrections has been carried out.

\section{Conclusions}
\label{conc}
In this paper, we have explored the effect of minimum length on BTZ black hole metrics by analyzing its thermodynamic properties.
We have implemented the minimum length contribution through a smeared mass distribution. In addition, we also introduce the minimum length through a collapsed shell mass density and a Lorentzian-like distribution with new terms being generated for the metric function.
With regard to the thermodynamic variables, we verified that the behavior of temperature, entropy and specific heat capacity indicates that there exists  a minimum mass for the black hole. We have found that the modified black hole is stable, and the specific heat capacity vanishes for $r_h=r_{min}$, which signalizes the presence
of remnants as the final stage of BTZ solution with minimum length. Moreover, in the situation where the mass is null, we have found non-zero results for thermodynamic quantities due to the minimum length effect.

\acknowledgments

We would like to thank CNPq, CAPES and CNPq/PRONEX/FAPESQ-PB (Grant nos. 165/2018 and 015/2019),
for partial financial support. MAA, FAB and EP acknowledges support from CNPq (Grant nos. 306398/2021-4, 312104/2018-9, 304290/2020-3).

\end{document}